\documentclass[aps,prl,twocolumn,showcaps,superscriptaddress,showpacs,amsmath,amssymb]{revtex4}

\usepackage{graphicx}% Include figure files
\usepackage{dcolumn}% Align table columns on decimal point
\usepackage{bm}% bold math
\usepackage[dvips]{epsfig}
\usepackage{color}
\usepackage{upgreek}

\newcommand{\unit}[1]{\;{\rm #1}}

\newcommand{\kB}{k_{\rm{B}}}
\newcommand{\kT}{k_{\rm{B}}T}
\newcommand{\xd}{\dot{x}}

\newcommand{\js}{j^{\rm{s}}}
\newcommand{\nus}{\nu^{\rm{s}}}
\newcommand{\ps}{\ensuremath{p^{\rm{s}}}}

\newcommand{\Df}{\Delta f}

\newcommand{\dx}{\partial_{x}}

\newcommand{\mean}[1]{\langle #1 \rangle}

\newcommand{\res}[1]{R_{#1}}
\newcommand{\Res}{\mathcal{R}}

\newcommand{\Cor}[1]{\mathcal{C}_{#1}}

\begin{document}

\title{Experimental Accessibility of Generalized Fluctuation-Dissipation Relations for Nonequilibrium Steady States}

\author{Jakob Mehl}
\affiliation{2. Physikalisches Institut, Universit\"at Stuttgart, Pfaffenwaldring 57, 70569 Stuttgart, Germany}
\author{Valentin Blickle}
\affiliation{2. Physikalisches Institut, Universit\"at Stuttgart, Pfaffenwaldring 57, 70569 Stuttgart, Germany}
\author{Udo Seifert}
\affiliation{II. Institut f\"ur Theoretische Physik, Universit\"at Stuttgart, Pfaffenwaldring 57, 70550 Stuttgart, Germany}
\author{Clemens Bechinger}
\affiliation{2. Physikalisches Institut, Universit\"at Stuttgart, Pfaffenwaldring 57, 70569 Stuttgart, Germany}
\affiliation{Max-Planck-Institut f\"ur Metallforschung, Heisenbergstrasse 3 ,70569 Stuttgart, Germany}

% ---------- Abstract ---------------------------------------------
\begin{abstract}

We study the fluctuation-dissipation theorem for a Brownian particle driven into a nonequilibrium steady state experimentally. We validate two different theoretical variants of a generalized fluctuation-dissipation theorem. Furthermore, we demonstrate that the choice of observables crucially affects the accuracy of determining the nonequilibrium response from steady state nonequilibrium fluctuations.

\end{abstract}


\maketitle

% ---------------------- Introduction -----------------------------------------

According to Onsager, in thermal equilibrium the reaction of a system to a small external perturbation and the decay of an internal fluctuation created by thermal noise are indistinguishable. This property, characteristic for the linear response around equilibrium, is expressed by the fluctuation-dissipation theorem (FDT)
\begin{equation}
 k_{\rm B}T\res{a,h}(t) = \mean{a(t)[-\partial_h\dot{E}(0)]}
 \label{eq:response-eq}
\end{equation}
which relates the time-dependent response $\res{a,h}(t)$ of an observable $a(t)$ to a perturbation $h$ to the correlation function between $a(t)$ and the derivative of the energy rate $\dot{E}$ with respect to $h$~\cite{kub91}. Here, $k_{\rm B}T$ is the thermal energy.

Since the FDT allows to determine response properties like mobilities or susceptibilities from equilibrium measurements of diffusivities or power spectra and vice versa, it has found widespread application in different scientific fields like statistical mechanics, biophysics, chemical or solid state physics~\cite{mar08}. In its original derivation the FDT holds only close to thermal equilibrium. Therefore, violations of the FDT are a clear fingerprint of a nonequilibrium system. So far, recent theoretical~\cite{spe06,che08,krue09,bai09,pro09,sei10} and experimental progress~\cite{gom09} demonstrated that the FDT can also be extended to a specific class of nonequilibrium systems, i.e., nonequilibrium steady states (NESSs). In~\cite{sei10} it has been shown that a similar expression as in Eq.~\eqref{eq:response-eq} can be obtained when the energy is replaced by the entropy within the correlation function on the right hand side. Furthermore, it was demonstrated that additional equivalent forms of the FDT exist, which in principle allows for infinitely many variants.

In this Brief Report, we experimentally demonstrate the validity of the FDT in a NESS for two different variants. Although both are equivalent from a theoretical point of view, large differences regarding the size of experimental errors exist. Therefore, the right choice of observables is important for the accurate determination of the response in such measurements.

% ---------------------- Setup ------------------------------------------------

% ---------------------- Figure 1 ---------------------------------------------
\begin{figure}
  \includegraphics[width=.9\linewidth]{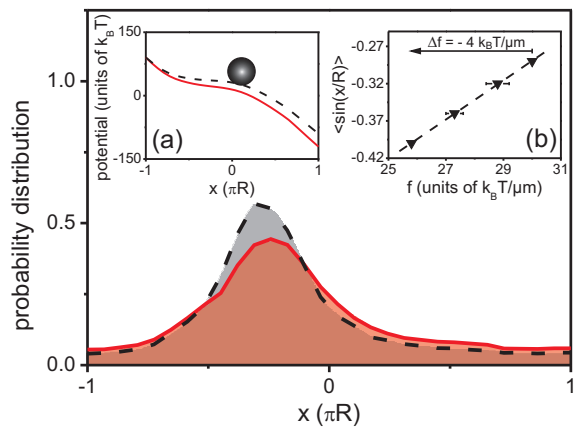}
\caption{(Color online) Normalized steady state probability distributions
$\ps_u(x)$ (unperturbed state; solid red line) and $\ps_p(x)$ (perturbed state; dashed black line). Inset: (a) Corresponding tilted potentials of unperturbed (red solid line) and perturbed (black dashed line) state. (b) $\mean{\sin(x/R)}$ for different driving forces. The black dashed line is a linear fit.\label{fig1}}
\end{figure}

Our experimental setup has been described in detail elsewhere~\cite{bli07a,bli09,fau95} and will be discussed here only briefly.
A colloidal silica particle of radius $r = 0.65\unit{\upmu m}$, immersed in water, is trapped within a three-dimensional torus of radius $R = 1.18\unit{\upmu m}$ by means of scanning laser tweezers. Using video microscopy, we track the angular coordinate $x$ of the particle with $-\pi R\leqslant x<\pi R$ with a spatial and temporal accuracy of $10\unit{nm}$ and $15\unit{ms}$, respectively. Since the torus is far away from the lower surface of the cuvette cell (approximately~$50\unit{\upmu m}$), hydrodynamic interactions with the sample cell are negligible~\cite{lutz06}. The tweezers scanning motion exerts a nonconservative force $f$ to the particle so that it circulates around the torus onto which an additional static sinusoidal potential $V(x)=(V_0/2)\sin(x/R)$ is imposed. The corresponding total force
\begin{equation}
 F(x) = -\partial_x V(x)+f,
\end{equation}
with $V_0=58\unit{\kT}$ and $f=30\unit{\kT/\upmu m}$~\cite{bli07b}, drives the particle into a NESS with an average circulation velocity of $v^{\rm s}=7\unit{\upmu m / s}$. In this NESS, the particle permanently dissipates heat into the surrounding media, while the system is still characterized by a time-independent probability distribution $\ps (x)$, measured as shown in Fig.~\ref{fig1}, and a constant current $\js$. Brownian particles driven by forces such as Eq. (2) have been discussed in a variety of fields as reviewed in~\cite{rei02}.

% ---------------------- Experiment -------------------------------------------

As in thermal equilibrium, the response of a NESS is defined as the
functional derivative of $\mean{a(t)}$ with respect to the perturbation $h$ in the limit of $h\rightarrow 0$: $\res{a,h}(t)\equiv\frac{\delta\mean{a(t)}}{\delta h(0)}$. However, in experiments such a functional derivative is not accessible. For a step-like perturbation of height $\Delta h$ integration of $\res{a,h}(t)$ leads to the integrated response
\begin{equation}
 \mathcal{R}(t)\equiv\int_0^t\res{a,h}(\tau)\,d\tau
 =\frac{\mean{a(t)}_u-\mean{a}_p}{-\Delta h},
\end{equation}
which is experimentally accessible. Here, the average $\mean{a(t)}_u$ is
determined after the perturbation is turned off and $\mean{a}_p$ is the average
of $a$ in the perturbed state, determined via a separate stationary
measurement. In principle, the FDT allows for many choices of the observable $a$ and the perturbation $h$. However, due to the symmetry of our system, a convenient choice for the observable is $a=\sin(x/R)$, which is obtained immediately by the measured angular coordinate $x$. For the perturbation we choose $\Delta h=\Df$, i.e., a small variation of the driving force $f$. In a dynamical experiment, we switch between the reference NESS and the perturbed NESS with a period of $6\unit{s}$ for approximately $4000$ times by instantaneously reducing the driving force $f$  by $\Delta f=-4\unit{\kT/\upmu m}$, while all other system parameters are kept constant. The linear dependence between $a$ and $f$, as shown in Fig.~\ref{fig1}(b), ensures that even for a force perturbation of $15\unit{\%}$ the system stays within the linear response regime and that $\Df=-4\unit{k_{\rm B}T/\upmu m}$ is sufficiently small.

% ---------------------- Figure 2 ---------------------------------------------
\begin{figure}
  \includegraphics[width=.9\linewidth]{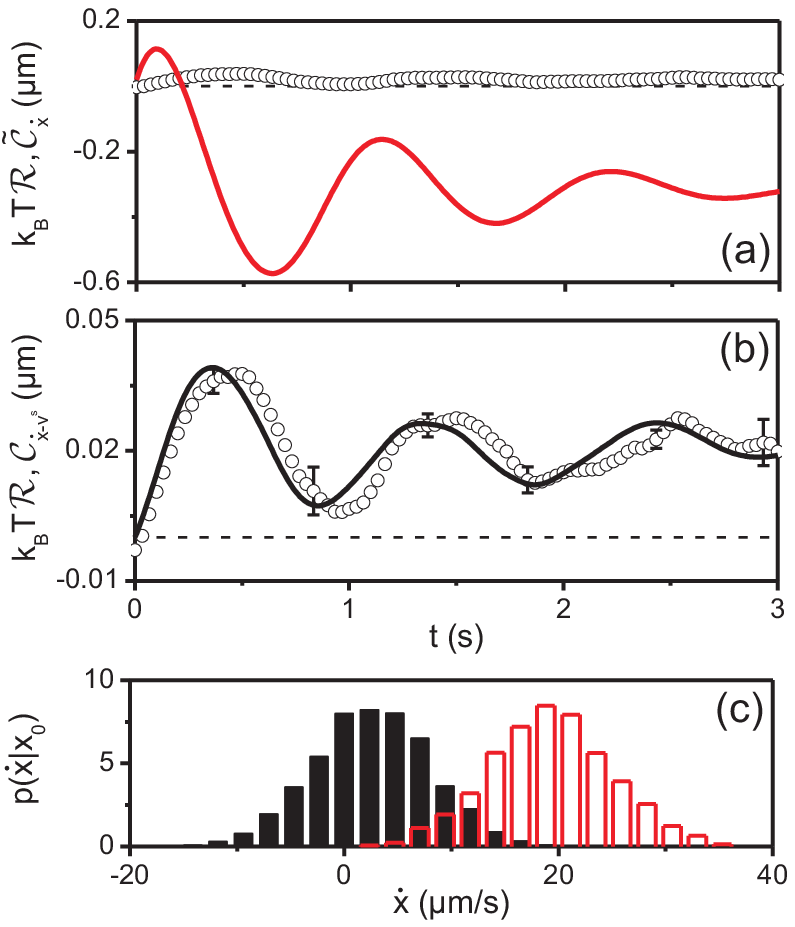}
\caption{(Color online) (a) Integrated response $k_{\rm B}T\,\Res(t)$ (open black circles) and integrated normalized correlation function
$\tilde{\mathcal{C}}_{\xd}(t)\equiv\Cor{\xd}(t)-\mean{a}\mean{\xd}t$ (red solid line). (b) $k_{\rm B}T\,\Res(t)$ (open black circles) and integrated correlation function $\Cor{\xd-\nus}(t)$ (black solid line) according to the integrated generalized FDT [Eq.~\eqref{eq:response-speck}]. (c) Conditional steady state probability distribution $p(\xd|x_0)$ at the flattest $x_0=-1.9\unit{\upmu m}$ (black closed bars; $\nus(x_0)=3\unit{\upmu m/ s}$, $F(x_0)=8\unit{\kT / \upmu m}$) and the steepest $x_0=-2.6\unit{\upmu m}$ (red open bars; $\nus(x_0)=20\unit{\upmu m/ s}$, $F(x_0)=55\unit{\kT / \upmu m}$) position in the tilted potential. \label{fig2}}
\end{figure}

First, we want to demonstrate the violation of the equilibrium FDT as given by Eq.~\eqref{eq:response-eq}. Since $x=\partial_{\Df}E$ is the conjugate observable to $\Df$ with respect to $E$, we get $k_{\rm B}T\, \res{a,\Df}(t)=\mean{a(t)\xd(0)}$ with $\xd$ the actual particle velocity. Thus, by defining for any observable $b$ the integrated correlation function as $\Cor{b}(t)\equiv\int_0^t\mean{a(\tau)b(0)}d\tau$, the integrated equillibrium FDT following from Eq.~\eqref{eq:response-eq} reads
\begin{equation}
 k_{\rm B}T\,\Res(t) = \Cor{\xd}(t).
 \label{eq:response-int}
\end{equation}
Figure~\ref{fig2}(a) shows $k_{\rm B}T\,\mathcal{R}(t)$ (open black circles), from a measurement involving the step-like perturbation, and the integrated normalized correlation funcion (red solid line), from a separate stationary measurement. Both quantities oscillate with a period of $1\unit{s}$ corresponding to the mean revolution time of the particle. This oscillatory behavior is an inherent feature of this NESS~\cite{bli09}, still present even though the motion of the particle is overdamped and inertia is completely negligible. In contrast, the response of an overdamped system at thermal equilibrium always decays exponentially. The fact that the response is more than one order of magnitude smaller than the correlation clearly demonstrates that the system is far away from thermal equilibrium.

% ---------------------- Theory -----------------------------------------------

The FDT can be generalized to a NESS by choosing the observable as conjugate to the perturbation $\Df$ with respect to the system entropy $s(t)=-k_{\rm{B}}\ln\ps(x(t))$~\cite{sei10},
\begin{equation}
  k_{\rm B}\,\res{a,\Df}(t)=\mean{a(t)[-\partial_{\Df}\dot{s}(0)]}.
 \label{eq:response-entropy}
\end{equation}
In thermal equilibrium Eq.~\eqref{eq:response-entropy} becomes Eq.~\eqref{eq:response-eq} since in the absence of an external driving force $\partial_{\Df}\dot{s}$ reduces to $\partial_{\Df}\dot{E}/T$ \cite{sei10}.
It is convenient to split the entropy production of the system into one of the medium $\dot{s}_{\rm{med}}$ and a total entropy production $\dot{s}_{\rm{tot}}$~\cite{sei05},
\begin{equation}
 \dot{s} = \dot{s}_{\rm{tot}}-\dot{s}_{\rm{med}}.
 \label{eq:entropy-add}
\end{equation}
According to Clausius, $\dot{s}_{\rm{med}}=\dot{Q}/T=\xd F/T$ corresponds to the heat flow $\dot{Q}$ into the thermal bath. The total entropy production rate is given by $\dot{s}_{\rm{tot}}=\xd \nus/(\mu_0T)$ with the bare mobility $\mu_0$ and the local mean velocity $\nus(x)\equiv\js/\ps(x)$, which corresponds to the average over all stochastic velocities for a fixed particle position $x$. After inserting Eq.~\eqref{eq:entropy-add} into Eq.~\eqref{eq:response-entropy} and integrating over time, the latter becomes~\cite{sei10,spe06}
\begin{equation}
 k_{\rm B}T\,\Res(t) = \Cor{\xd}(t)-\Cor{\nus}(t) \equiv 
 \Cor{\xd-\nus}(t).
 \label{eq:response-speck}
\end{equation}
Here, the first correlation function corresponds to the one occurring also in the integrated equilibrium FDT [Eq.~\eqref{eq:response-int}] but now evaluated under nonequilibrium conditions. The second term reflects the violation of detailed balance, which vanishes under equilbrium conditions ($\nus\rightarrow0$).

% ---------------------- Experiment -------------------------------------------

Our measurements [see Fig.~\ref{fig2}(b)] confirm the generalized FDT from Eq.~\eqref{eq:response-speck} connecting response and correlation function since within the statistical errors $k_{\rm B}T\,\mathcal{R}(t)$ (open black circles) and $\Cor{\xd-\nus}(t)$ (black solid line) coincide. We want to emphasize that Eq.~\eqref{eq:response-speck} was tested here over more than two relaxation times ($\tau_r=1.4\unit{s}$) corresponding to about three particle revolutions. Due to this short relaxation time (in our experiment $\tau_r\sim R^2/D_0$ with $D_0=\kB T/(6\pi\eta r)$ the free diffusion coefficient and $\eta$ the viscosity of the solvent~\cite{bli09}), in contrast to previous work~\cite{gom09}, we were able to observe the generalized FDT along the full damped oscillations of $k_{\rm B}T\,\mathcal{R}(t)$.

The FDT in the NESS can be brought into a form even more reminiscent to the equilibrium FDT by introducing the velocity $v(t)\equiv\xd(t)-\nus(x(t))$, which is measured with respect to the local mean velocity $\nus$~\cite{spe06,che08,gom09}. Equation~\eqref{eq:response-speck} then reads $k_{\rm B}T\,\Res(t) =\Cor{v}(t)$. A physical reason for this restoration of the equilibrium FDT in the locally comoving frame can be seen in measuring the conditional velocity distributions $p(\xd|x_0)$ in this frame. Their shapes are the same at any position $x_0$ as illustrated in Fig.~\ref{fig2}(c) for two specific positions. Both distributions resemble Gaussians with a mean equals to the local mean velocity $\nus(x_0)$ and a width of $\sigma^2=2D_0/\Delta t$, where $\Delta t=20\unit{ms}$ is the time interval over which the velocity $\xd$ was determined. Thus, locally the velocity fluctuations are indistinguishable from equilibrium ones even though nontrivial velocity correlations develop in a NESS even when expressed in the locally comoving frame.

% ---------------------- Figure 3 ---------------------------------------------
\begin{figure}
  \includegraphics[width=.9\linewidth]{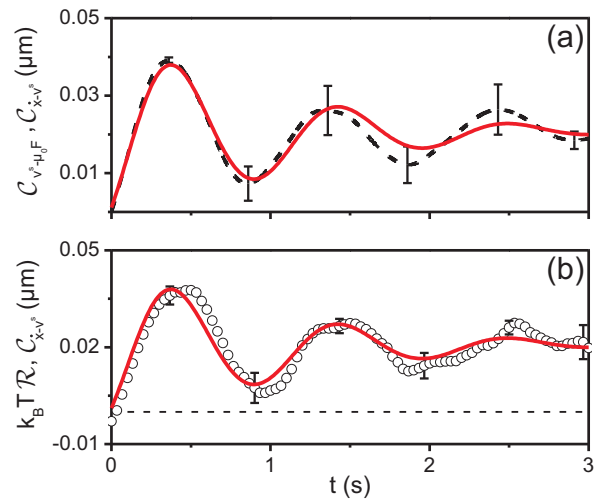}
\caption{(Color online)  (a) Integrated correlation functions
$\Cor{\xd-\nus}(t)$ (black dashed line) and $\Cor{\nus-\mu_0F}(t)$ (red solid line) indicating different forms of the generalized FDT. (b) Comparison between $k_{\rm B}T\,\Res(t)$ (black open circles) and $\Cor{\nus-\mu_0F}(t)$ (red solid line). \label{fig3}}
\end{figure}

% ---------------------- Theory -----------------------------------------------

Somewhat surprisingly, the conjugate observable appearing in the correlation function on the right hand side of the generalized FDT in Eq.~\eqref{eq:response-entropy} is not unique as observed in~\cite{sei10}, where a classification of the different variants was suggested. This equivalence in a theoretical perspective, however, does not extend to an equivalence in terms of experimental accessibility as we next demonstrate. The earliest variant of the FDT in a NESS is based on Agarwal~\cite{aga72} and reads in our notation
\begin{equation}
 k_{\rm B}T\,\res{a,\Df}(t) = \mean{a(t)[-D_0\dx\ln\ps(x(0))]}.
 \label{eq:response-agarwal}
\end{equation}
Here, the conjugate observable is derived from the steady state distribution which leads to an observable in the state (i.e., configuration) space that does not involve a time derivative. Consequently, it is more easily accessible experimentally than the strongly fluctuating stochastic velocity $\xd$ appearing in Eq.~\eqref{eq:response-speck}.

In a steady state, the constant probability current can be written as $\js=\mu_0F(x)\ps(x)-D_0\dx\ps(x)$~\cite{ris96}. With the definition of the local mean velocity, we obtain
\begin{equation}
 -D_0\dx\ln\ps(x) = \nus(x)-\mu_0F(x).
 \label{eq:gradient}
\end{equation}
Inserted into Eq.~\eqref{eq:response-agarwal} and integrated over time, we get
\begin{equation}
 k_{\rm B}T\,\Res(t) = \Cor{\nus}(t)-\Cor{\mu_0F}(t) \equiv 
 \Cor{\nus-\mu_0F}(t).
 \label{eq:corr}
\end{equation}
Since both the integrated generalized FDT from Eqs.~\eqref{eq:response-speck} and \eqref{eq:corr} are valid for any steady state, their correlation functions must be identical
\begin{equation}
 \Cor{\xd-\nus}(t) = \Cor{\nus-\mu_0F}(t).
\end{equation}
Even though their second arguments are different, there is indeed a very good agreement between these correlation functions, as seen in Fig. \ref{fig3}(a). The deviation, which occurs only for times larger than $1.5\unit{s}$, can be attributed to statistical errors which increase in time. Compared to the previously shown function $\Cor{\xd-\nus}(t)$ [see Fig.~\ref{fig2}(b)] the correlation function $\Cor{\nus-\mu_0F}(t)$ [see Fig.~\ref{fig3}(b)] traces the integrated response even better. As most pronounced in the time interval between $1.5$ and $3\unit{s}$, the phase and the amplitude of the oscillations are represented by $\Cor{\nus-\mu_0F}(t)$ more precisely.

% ---------------------- Figure 4 ---------------------------------------------
\begin{figure}[t]
  \includegraphics[width=.9\linewidth]{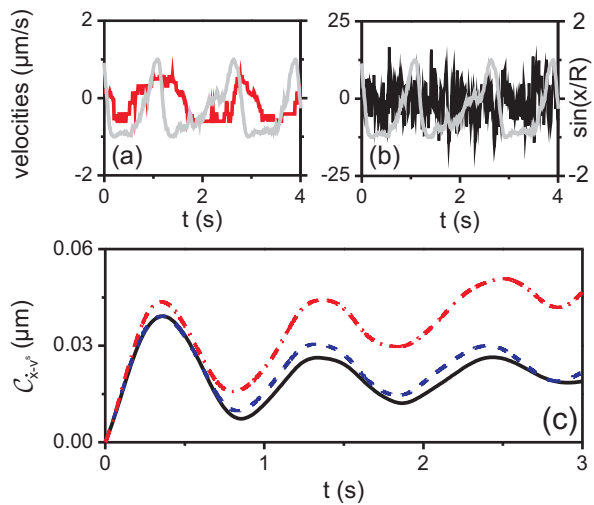}
\caption{(Color online)  (a), (b) Quantities involved in the correlation
functions: $\sin(x/R)$ (light gray line), $\nus-\mu_0 F$ (red line) and $\xd-\nus$ (black line). (c) Integrated correlation function $\Cor{\xd-\nus}(t)$ determined from different raw data lengths. The black line corresponds to the full data set of $3000\unit{s}$ whereas for the blue dashed (red dashed-dotted) line only one half (quarter) of the original data was used. \label{fig4}}
\end{figure}

To understand why $\Cor{\nus-\mu_0F}(t)$ leads to a more accurate
determination of $k_{\rm B}T\,\Res(t)$ we next concentrate on the terms involved in these two different correlation functions. Figures~\ref{fig4}(a) and \ref{fig4}(b) show their time evolution along the trajectories over a time interval of $3\unit{s}$. Although leading to the same correlation function when correlated with $a=\sin(x/R)$ (light gray line), the trajectories of $\nus-\mu_0 F$ (red line) and $\xd-\nus$ (black line) are different. While the first trajectory is rather smooth, in the second one, due to the appearance of $\xd$, strong fluctuations are present. The amplitude of these fluctuations is more than ten times larger than the maximal velocity variation of $\nus-\mu_0F$. Since for a reliable average $\mean{\dots}$ many realizations have to be sampled, it is evident that the correlation $\Cor{\xd-\nus}(t)$ cannot be determined with the same statistical precision than $\Cor{\nus-\mu_0F}(t)$. This observation is supported by Fig.~\ref{fig4}(c) where, in order to examine the impact of statistics, only parts of the acquired raw data were evaluated. The black line shows $\Cor{\xd-\nus}(t)$ calculated from the complete raw trajectory of length $3000\unit{s}$. For the blue and the red line the length of the trajectory was reduced by a factor of two and four, respectively. In the latter case the strong deviation to the black curve is obvious. In contrast, $\Cor{\nus-\mu_0F}(t)$ is almost insensitive to a reduction of the statistics. Indeed, reducing the length of the evaluated data to $300\unit{s}$ (one order of magnitude) does not influence the shape of $\Cor{\nus-\mu_0F}(t)$ visibly.

The insight gained in this model system can be transferred to other systems as well: If one wants to obtain the response of a NESS from a measurement of its stationary correlations, one should use the variant of the generalized FDT not involving time derivatives for its better statistical convergence properties. In this respect, the situation in experiments is somewhat different from that in simulations, where the correlation between $a(t)$ and the {\it a priori} known Langevin noise $(\xd-\mu_0F)/2=\xi/2$ determines the response most conveniently~\cite{sei10}.

% ---------------------- Summary ----------------------------------------------

In summary, we studied the FDT for a driven Brownian particle experimentally.
For this paradigmatic system, a kind of "Ising-Model" of nonequilibrium steady states, we validated two different generalizations of the FDT to the nonequilibrium regime. We demonstrated that the right choice of observables affects the errors when calculating the response from a measurement of stationary correlations.

We thank B. Lander and T. Speck for stimulating discussions and suggestions.
V.B. (BL 1067) and U.S. (SE 1119/3) were supported by the Deutsche Forschungsgemeinschaft.

% ---------------------- Bibliography -----------------------------------------

\end{document}